\documentclass[pra,twocolumn,epsfig,rotate,showpacs]{revtex4}
\usepackage{amsfonts} 
\usepackage{amsmath}
\usepackage{amssymb}
\usepackage{bm}
\usepackage[usenames, dvipsnames]{color} 
\usepackage{dcolumn}
\usepackage{epic} 
\usepackage{epsfig}
\usepackage{graphicx}
\usepackage[abs]{overpic}
\usepackage[lofdepth,lotdepth,caption=false]{subfig}

\usepackage[breaklinks,colorlinks = true,linkcolor = blue,urlcolor=blue,citecolor=blue]{hyperref}
\usepackage{soul}
\usepackage{times}
\usepackage{ulem}
\usepackage{wrapfig} 
\usepackage{xy} 

\newcommand{\beq}{\begin{eqnarray}}
\newcommand{\eeq}{\end{eqnarray}}
\begin{document}
\title{Non-Trivial Excited State Coherence Due to Two Uncorrelated Partially Coherent Fields}
\author{Z.S. Sadeq}
\affiliation{Department of Physics, 60 St. George Street
University of Toronto, Toronto, M5S 1A7, Canada}
\date{\today}
\begin{abstract}
We analyze a model where a closed $V$ system is excited by two uncorrelated partially coherent fields. We use a collisionally broadened CW laser, which is a good model for an experimentally realizable partially coherent field, and show that it is possible to generate excited state coherences even if the two fields are uncorrelated. This transient coherence can be increased if splitting between the excited states is reduced relative to the radiation coherence time, $\tau_{d}$. For small excited state splitting, one can use this scheme to generate a long lived coherent response in the system. 
\end{abstract}
\email{sadeqz@physics.utoronto.ca}
\maketitle

\section{Introduction}
Generating coherences from incoherent sources has been discussed extensively in literature \cite{aharony,kozlov, scully, brumer, mancal,plenio}. It has been established that one can generate transient coherences even from incoherent sources \cite{aharony,kozlov,scully,plenio,tors, timur2}. Various investigations of incoherent pumping in three level atoms \cite{scully, kozlov} has already been done. These reports have shown that it is indeed possible to use an incoherent field to induce transient coherent dynamics between states. Using this transient coherence, they have demonstrated the ability to do lasing without inversion \cite{scully}, quenching of spontaneous emission \cite{zub}, all coherent phenomena induced by incoherent pumps. The long time result of interaction of matter with incoherent light has been investigated and is found to be a mixed state \cite{brumer}. However the problem with using white noise is its coherence properties. Firstly, white noise is not realizable in the laboratory and is a mathematical concept rather than a physical one. Second, it is very difficult to compute the first order coherence function. 

However, these approaches (\cite{scully, kozlov}) are not truly incoherent in the sense that they use the same field to induce transitions from a common ground state $|g\rangle$ to excited states $|1\rangle$ and $|2\rangle$ as well as the fact that they use white noise, a source with a poorly defined first order coherence function $g^{(1)}(\tau)$.

Here we demonstrate how, when the correlation the field shares with itself and the atom at a time $t_{0}$ is removed, white noise cannot generate coherences between excited states in a $V$ configuration. Instead, we advocate the use of a collisionally broadened
CW source \cite{loudon}. This is given by the two time correlation function:

\begin{equation}
\langle\epsilon(t')\epsilon^{*}(t'')\rangle=\epsilon_{0}^{2}e^{-i\omega_{0}\left(t'-t''\right)}e^{-\frac{|t'-t''|}{\tau_{d}}}
\label{corr1}
\end{equation}

In this model, $\omega_{0}$ represents the frequency center of the radiation, and $\epsilon_{0}^{2}$ represents the field intensity or electric field strength.  The coherence time of this
radiation is given by $\tau_{d}= \hbar / kT$ where $\hbar$ is the reduced Planck
constant, $k$ is Boltzmann's constant, and $T$ is temperature \cite{wolf}.


The first order coherence function for a collisionally broadened CW source is much easier to compute \cite{loudon} than that of white noise. It is expressed as a function of the difference between the two times $\tau \equiv t^{'} - t^{''}$. 

\begin{equation}
g^{(1)}(\tau) = \text{exp}(-i\omega_{0}\tau - \frac{|\tau|}{\tau_{d}})
\label{g1wien}
\end{equation}

From Eq. \ref{g1wien} it is evident that the source is completely coherent for $\tau = 0$. This coherence eventually exponentially decays as a function of $\tau$ scaled by a characteristic radiation coherence time, $\tau_{d}$. This is a much more realistic noisy source than white light. The first order coherence function has a finite level of coherence at $\tau = 0$, as opposed to white noise which does not. In fact, at $\tau = 0$,  the first order correlation function of white noise has a singularity.

In this paper we use a collisionally broadened CW source, each tuned to the transition between the ground and excited states in a closed but not degenerate $V$ configuration. These two noisy lasers are also forced to be uncorrelated with each other. We demonstrate that the two excited states demonstrate a transient coherence associated with this noisy electric field.

This paper is organized as follows: Section II discusses the excitation of a $V$ system with two uncorrelated white noise sources, Section III discusses the irradiation of a $V$ level system with two uncorrelated collisionally broadened CW sources and the paper is concluded in Section IV.  


\section{\label{app:whitenoise}$V$ level atom subject to two white noise electric fields}

An example of incoherent pumping is a closed $V$ level atom irradiated by two white noise electric fields similar to the approach of \cite{kozlov, scully}. It is described by the Hamiltonian outlined in Eq. (3). Here we make the rotating wave (RWA) and electric dipole approximation (EDA). This describes an interaction between a three level system and two electric fields that couple the following transitions $|g\rangle \rightarrow |1\rangle$ and $|g\rangle \rightarrow |2\rangle$. $\mu_{1}$ is the dipole moment of the $|g\rangle \rightarrow |1\rangle$ transition and $\mu_{2}$ is the dipole moment of the $|g\rangle \rightarrow |2\rangle$ transition.
In the interaction picture, the Hamiltonian is of the form:

\begin{equation}
\mathcal{H}_{I}=\left(\begin{array}{ccc}
0 & e^{-i\omega_{1g}t}\overline{E_{1}(t)} & e^{-i\omega_{2g}t}\overline{E_{2}(t)}\\
e^{i\omega_{1g}t}\overline{E_{1}^{*}(t)} & 0 & 0\\
e^{i\omega_{2g}t}\overline{E_{2}^{*}(t)} & 0 & 0
\end{array}\right)
\end{equation}

\begin{figure}[b]
\begin{center}
\includegraphics[scale=0.48]{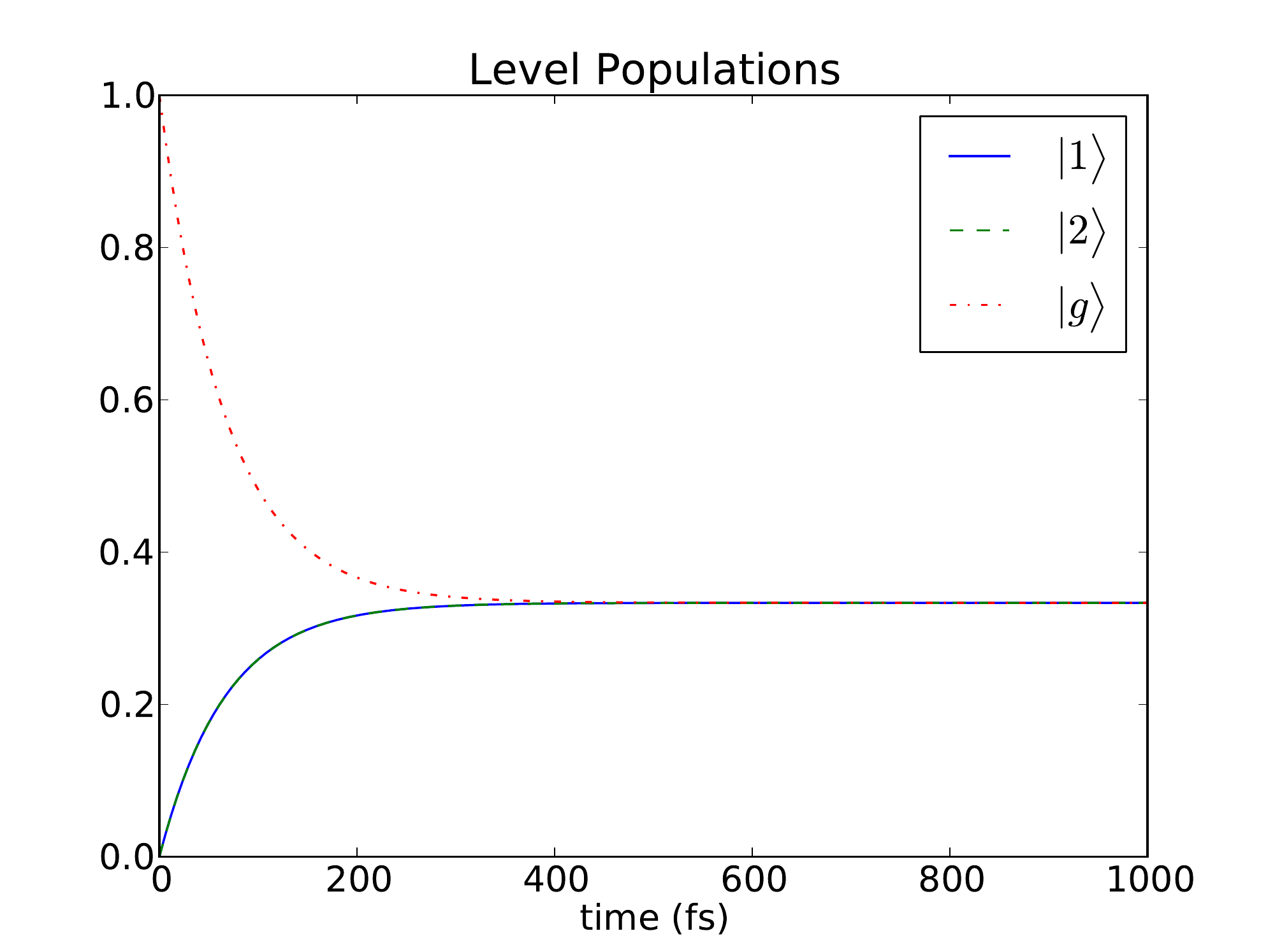}
\caption{(Color online) Plot of excited and ground state populations of the model $V$ system. In the long time limit, it can be seen that the populations equilibriate to $1/3$. Pump power to both levels was set to $\mu_{i}\mathcal{R}_{i} / \hbar = 250$ THz. }
\label{whitenoiseres}
\end{center}
\end{figure}

\begin{figure}[b]
\begin{center}
\includegraphics[scale=0.48]{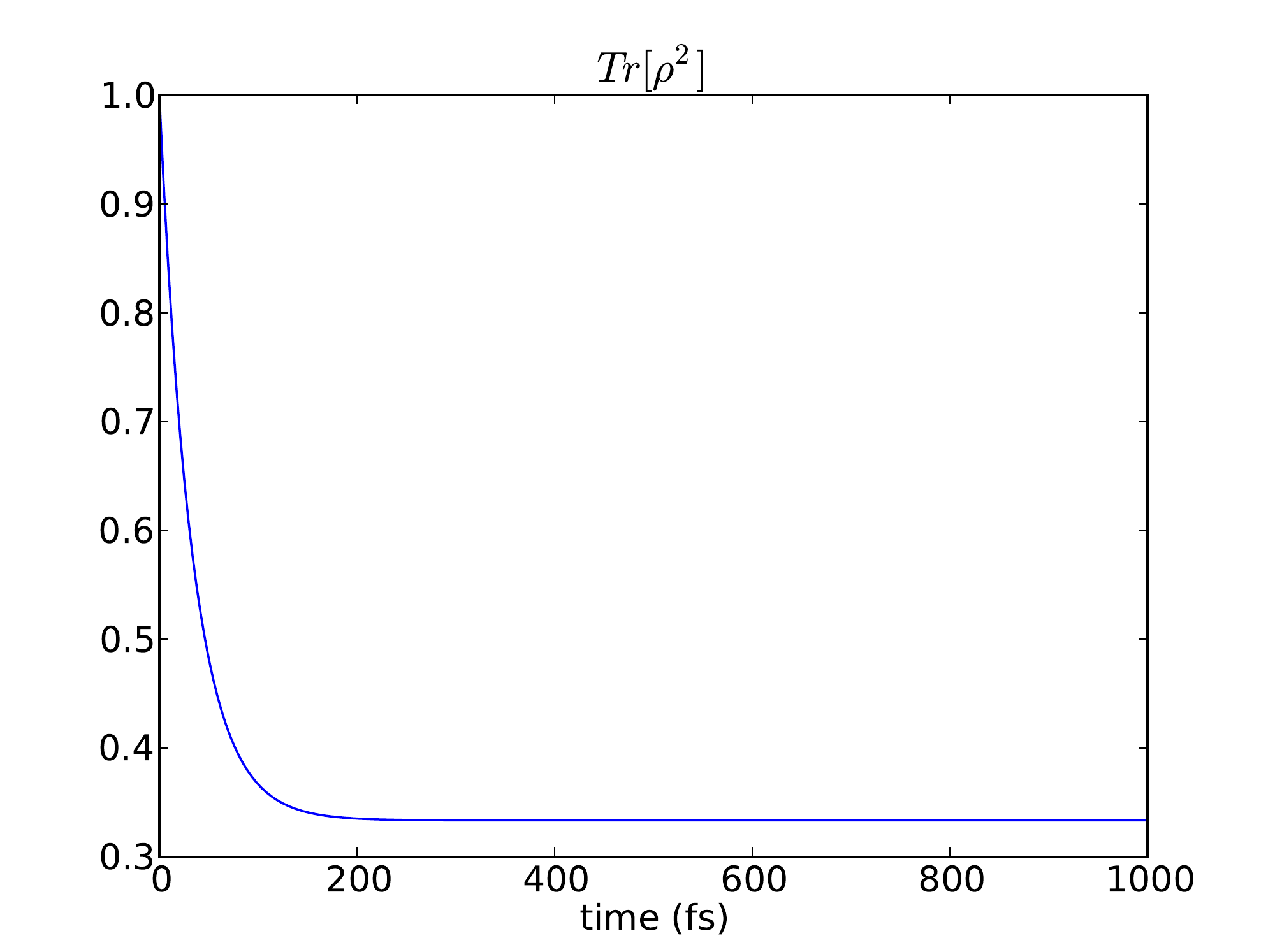}
\caption{Plot of the purity $\text{Tr}[\rho^{2}]$ of the $V$ system. From the plot of the purity, it is clear that the state quickly becomes a mixed state with the purity reaching values of $1/3$ quickly. Pump power to both levels was set to $\mu_{i}\mathcal{R}_{i} / \hbar = 250$ THz.}
\label{whitenoiseres2}
\end{center}
\end{figure}

Here we define: $\overline{E_{i}(t)}\equiv-\mu_{i}\epsilon_{i}(t)$. The white noise statistics of the electric fields are given by:

\begin{equation}
\langle \epsilon^{*}_{i}(t) \epsilon_{j}(t') \rangle = \delta_{ij}\mathcal{R}_{i} \delta(t-t')
\label{whitenoisecorr}
\end{equation}

The laser power is represented by the parameter $\mathcal{R}_{i}$. We acquire a solution by formally integrating Eq. (5):
\begin{equation}
\dot{\rho}=-\frac{i}{\hbar}\left[\mathcal{H}_{I}(t),\rho(0)\right]-\frac{1}{\hbar^{2}}\int dt'\left[\mathcal{H}_{I}(t),\left[\mathcal{H}_{I}(t'),\rho(t')\right]\right]
\end{equation}

After formal integration and ensemble averaging over the field. This leads to the following equation of motions for the populations of the states (in the Schroedinger picture):

\begin{equation}
\dot{\rho}_{gg} = \frac{2\mathcal{R}_{1}\mu^{2}_{1}}{\hbar^{2}}\rho_{11}(t) + \frac{2\mathcal{R}_{2}\mu^{2}_{2}}{\hbar^{2}}\rho_{22}(t) - \frac{2\rho_{gg}(t)}{\hbar^{2}}(\mathcal{R}_{1}\mu^{2}_{1} + \mathcal{R}_{2}\mu^{2}_{2})
\end{equation}

\begin{equation}
\dot{\rho}_{11} = \frac{2\mu^{2}_{1}\mathcal{R}_{1}}{\hbar^{2}}\left( \rho_{gg}(t) - \rho_{11}(t) \right)
\end{equation}

\begin{equation}
\dot{\rho}_{22} = \frac{2\mu^{2}_{2}\mathcal{R}_{2}}{\hbar^{2}}\left( \rho_{gg}(t) - \rho_{22}(t) \right)
\end{equation}

The equation of motion of the coherences between the two excited states is given by:

\begin{equation}
\dot{\rho}_{12}(t) =-\frac{i}{\hbar}\omega_{12}\rho_{12}(t) - \frac{1}{\hbar^2}( \mu^{2}_{1}\mathcal{R}_{1} \rho_{12}(t) + \mu^{2}_{2}\mathcal{R}_{2} \rho_{12}(t)) 
\end{equation}

We numerically solve the above equations for an initial condition of $\rho(0) = |g\rangle\langle g|$. These results are presented in Fig \ref{whitenoiseres}. From these calculations, it is apparent that it is not possible to generate coherences from these two uncorrelated white noise lasers. In fact, in the steady state all the populations equilibrate, i.e. $\rho_{gg} = \rho_{11} = \rho_{22} = 1/3$. Purity of the system as well as excited state coherence fraction, $\mathcal{C} \equiv |\rho_{12}|/(\rho_{11} + \rho_{22})$ is also plotted and presented in Fig \ref{whitenoiseres2}.

This scenario leads to the creation of a mixed state without ever generating coherences between the excited states. This is due to the Kronecker delta in the correlation function of the white noise (Eq (\ref{whitenoisecorr})). The Kronecker delta ensures that the two fields inducing the transitions to the upper levels are not correlated with each other at any time. In literature schemes \cite{kozlov, scully}, this condition is not enforced. As a result, the field is always correlated with itself in time. By eliminating this self correlation, one eliminates all transient coherence that might appear because of it. Removing this correlation is important as the $g^{(1)}(\tau)$ of white noise encounters a singularity at $\tau = 0$ and it is this singularity that drives the atom to produce coherences between the excited states.

Our approach should be contrasted to the work of ref \cite{kozlov, scully} which utilizes this initial self correlation to induce coherences between radiation uncoupled states in both the $\Lambda$ and $V$ architectures. The scenario presented in this paper is unique because it removes this initial self correlation. By removing this self correlation, one cannot induce excited state coherences using white noise.

\section{$V$ Level System Subject to Two Uncorrelated Collisionally Broadened CW Lasers}

\begin{figure}[h]
\begin{center}
\includegraphics[scale=0.45]{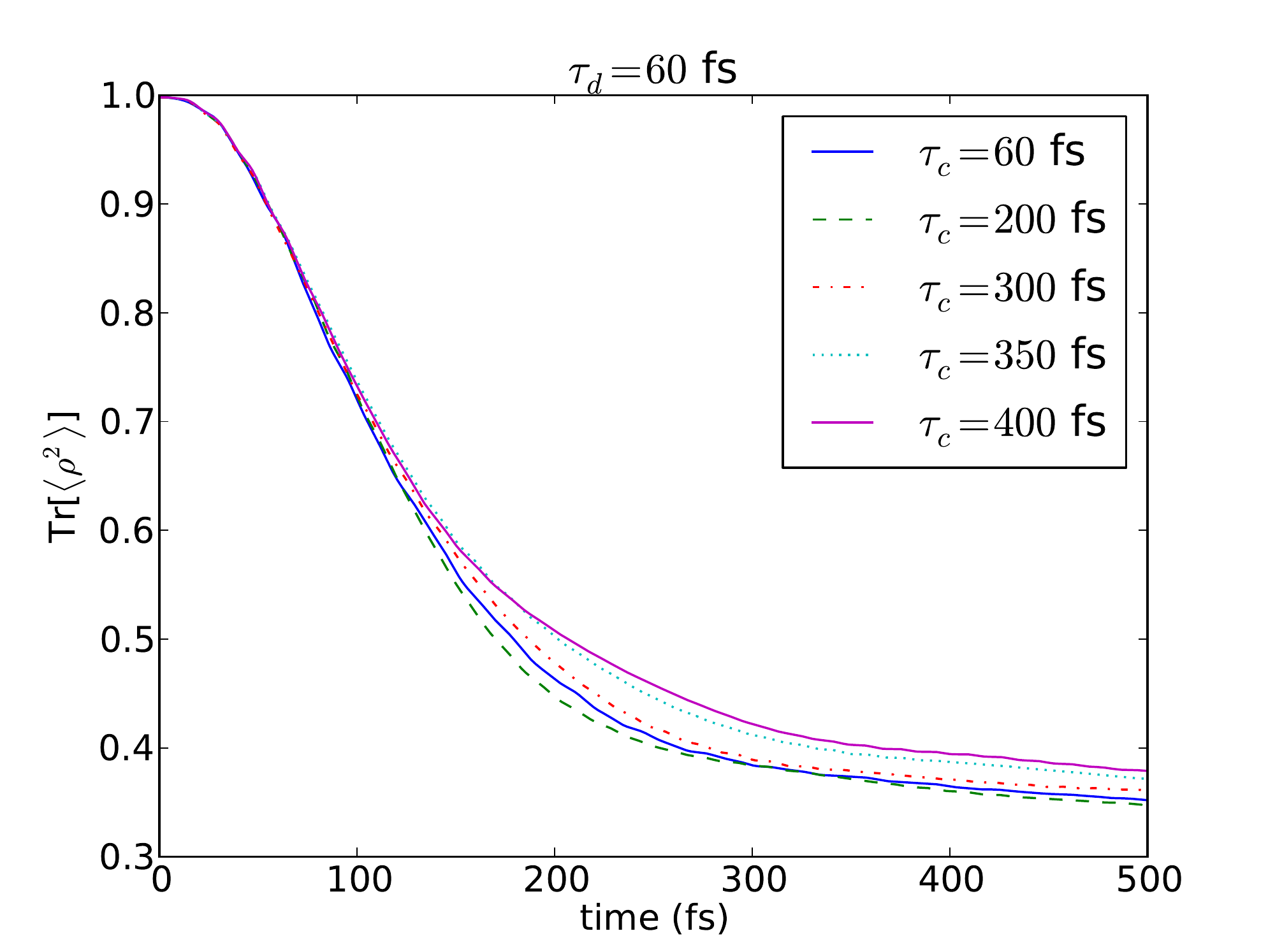}
\includegraphics[scale=0.45]{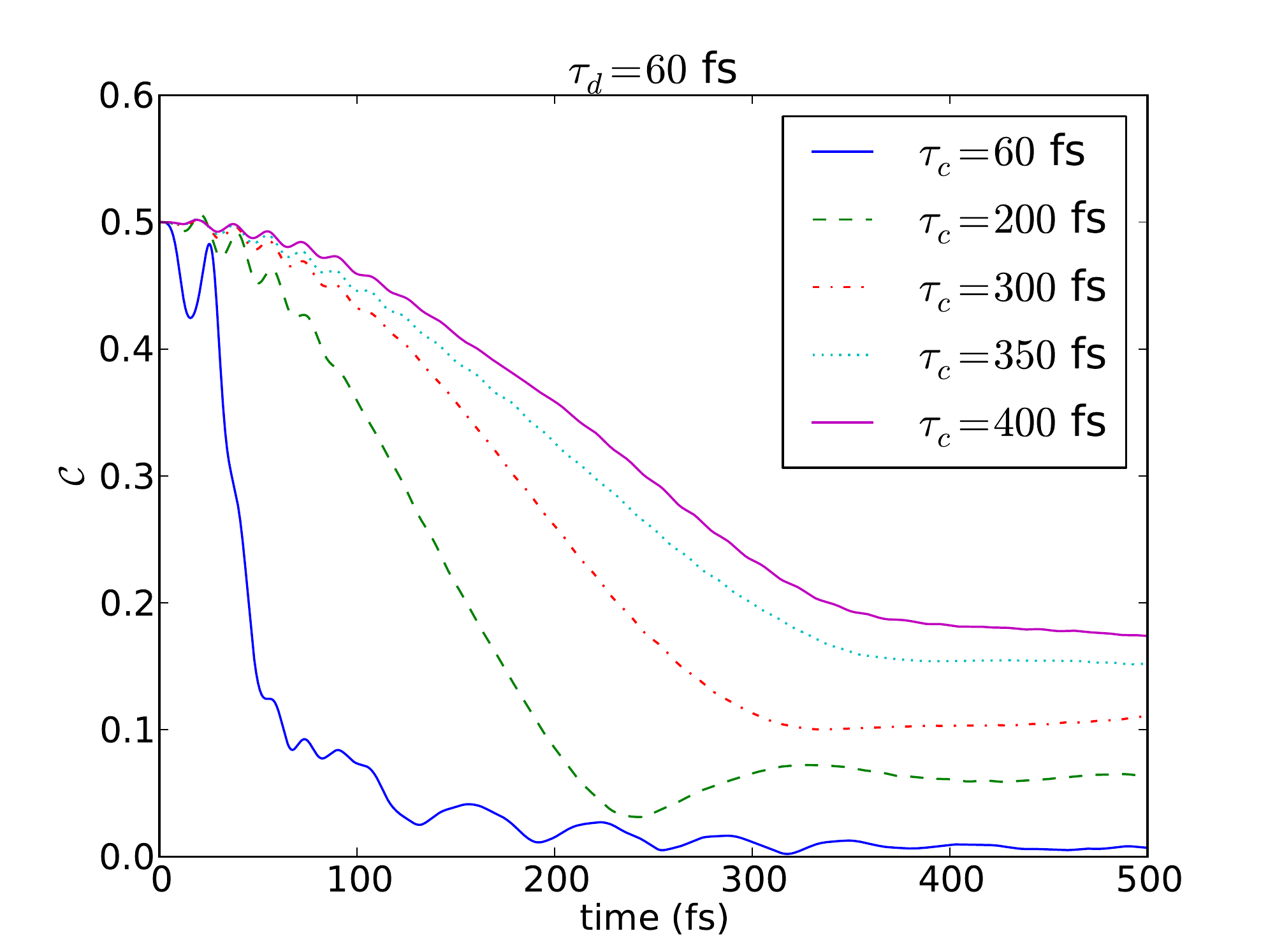}
\caption{(Color online)  Purity, Tr[$\rho^{2}$]  (top), for various excited state spacings is plotted as a function of time. We define $\tau_{c} = \frac{2\pi}{\omega_{21}}$ as the characteristic excited state period (bottom) excited state coherences as a fraction of excited state population $\mathcal{C} \equiv |\rho_{12}|/(\rho_{11} + \rho_{22})$  is plotted for various excited state spacings. It is evident that as excited state splitting decreases (i.e. $\tau_{c}$ increases) the system becomes increasingly more coherent. The radiation coherence time, $\tau_{d}=$60 fs for all figures.}
\label{wienerres}
\end{center}
\end{figure}

\begin{figure}[b]
\begin{center}
\includegraphics[scale=0.45]{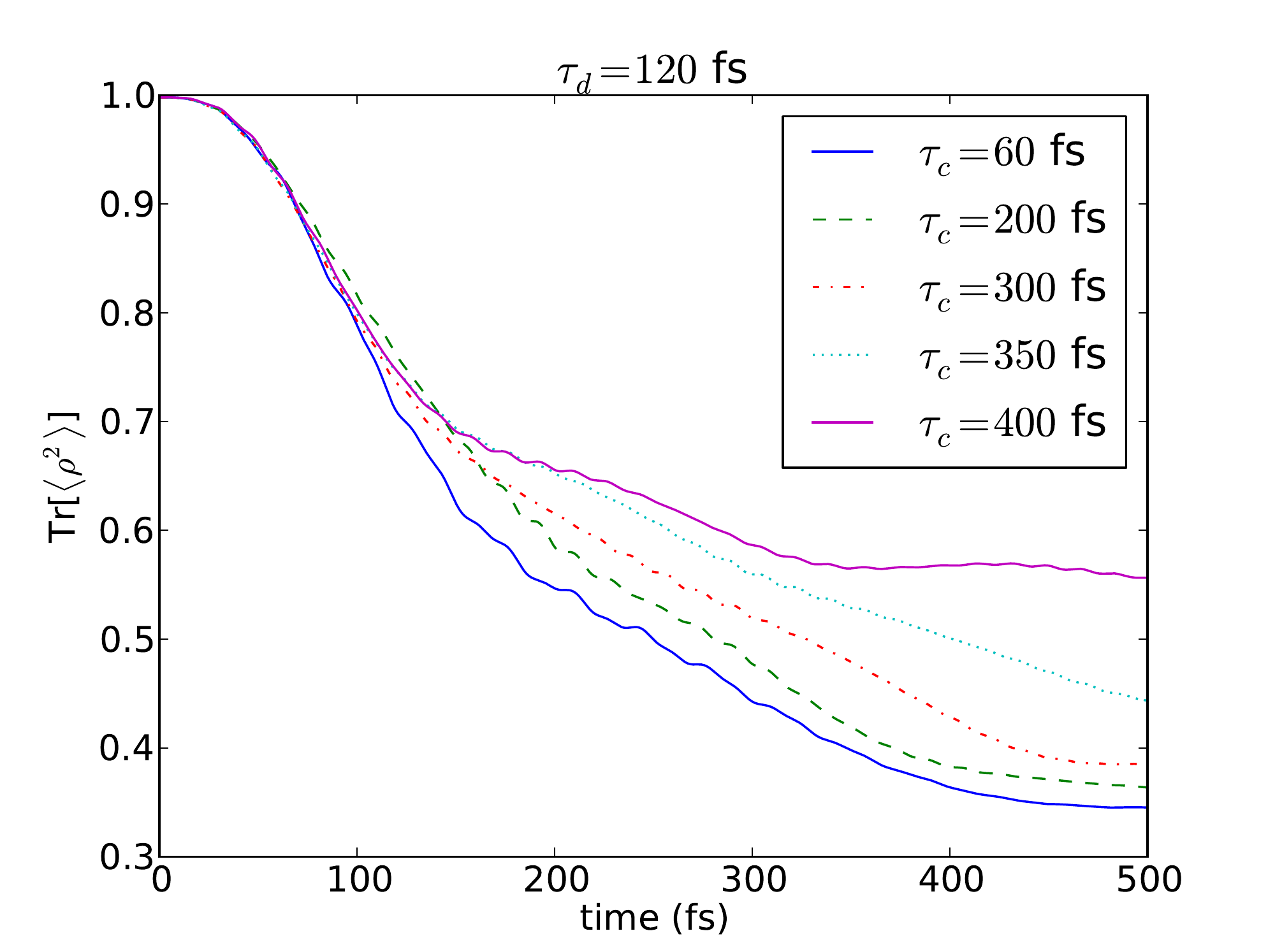}
\includegraphics[scale=0.4]{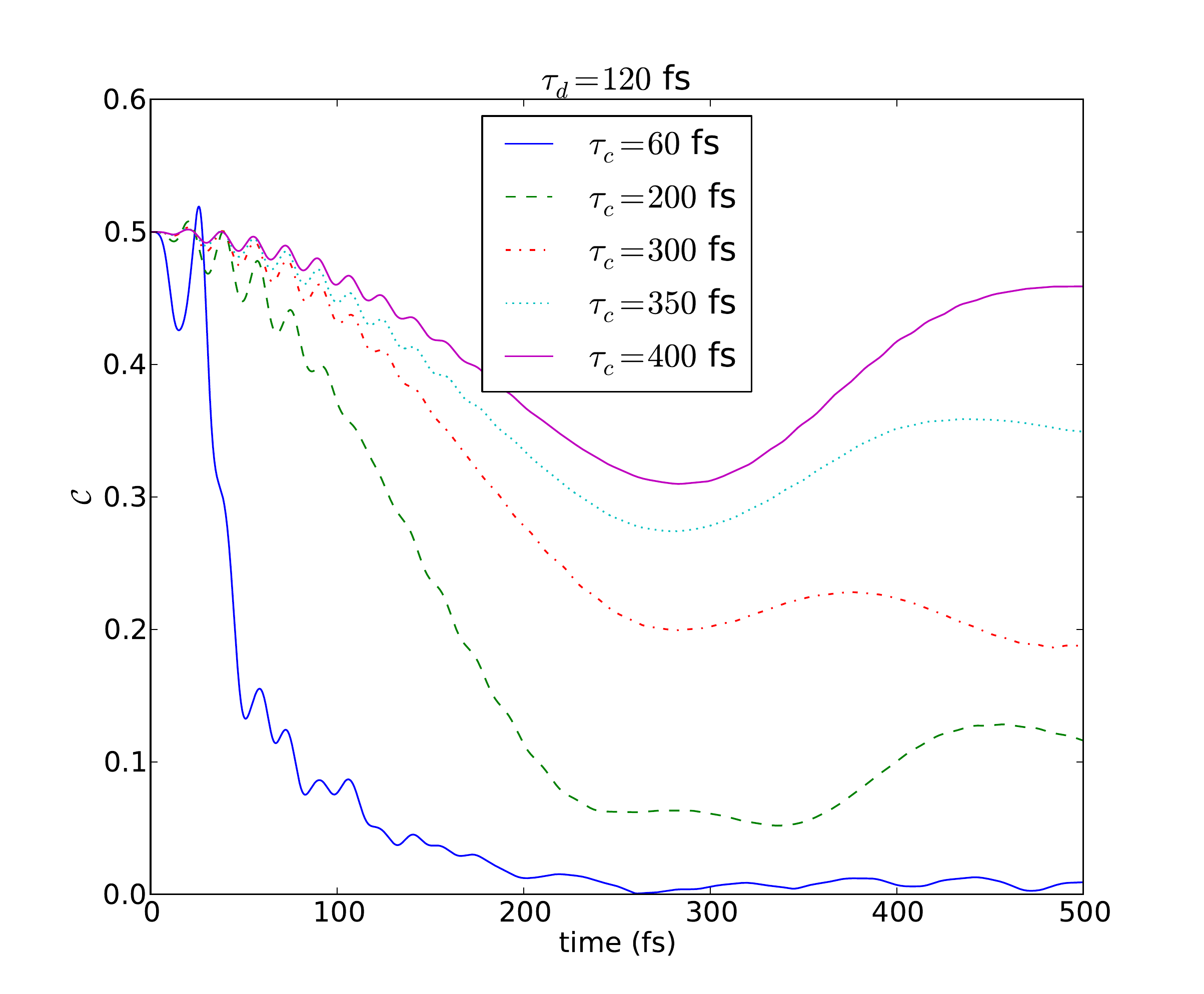}
\caption{(Color online) Purity, Tr[$\rho^{2}$] (top) , is plotted for various excited state spacings. We define $\tau_{c} = \frac{2\pi}{\omega_{21}}$ to be the characteristic excited state period (bottom) excited state coherences as a fraction of excited state population ($\mathcal{C} \equiv |\rho_{12}|/(\rho_{11} + \rho_{22})$) for various excited state spacings plotted as a function of time. In the case of $\tau_{c} = 400$ fs the excited state subspace remains unusually coherent with $\mathcal{C}$ approaching $\approx 0.45$ which is similar to a coherent superposition. The radiation coherence time, $\tau_{d}=120$ fs for all figures. }
\label{wienerres2}
\end{center}
\end{figure}

\begin{figure}[h]
\begin{center}
\includegraphics[scale=0.45]{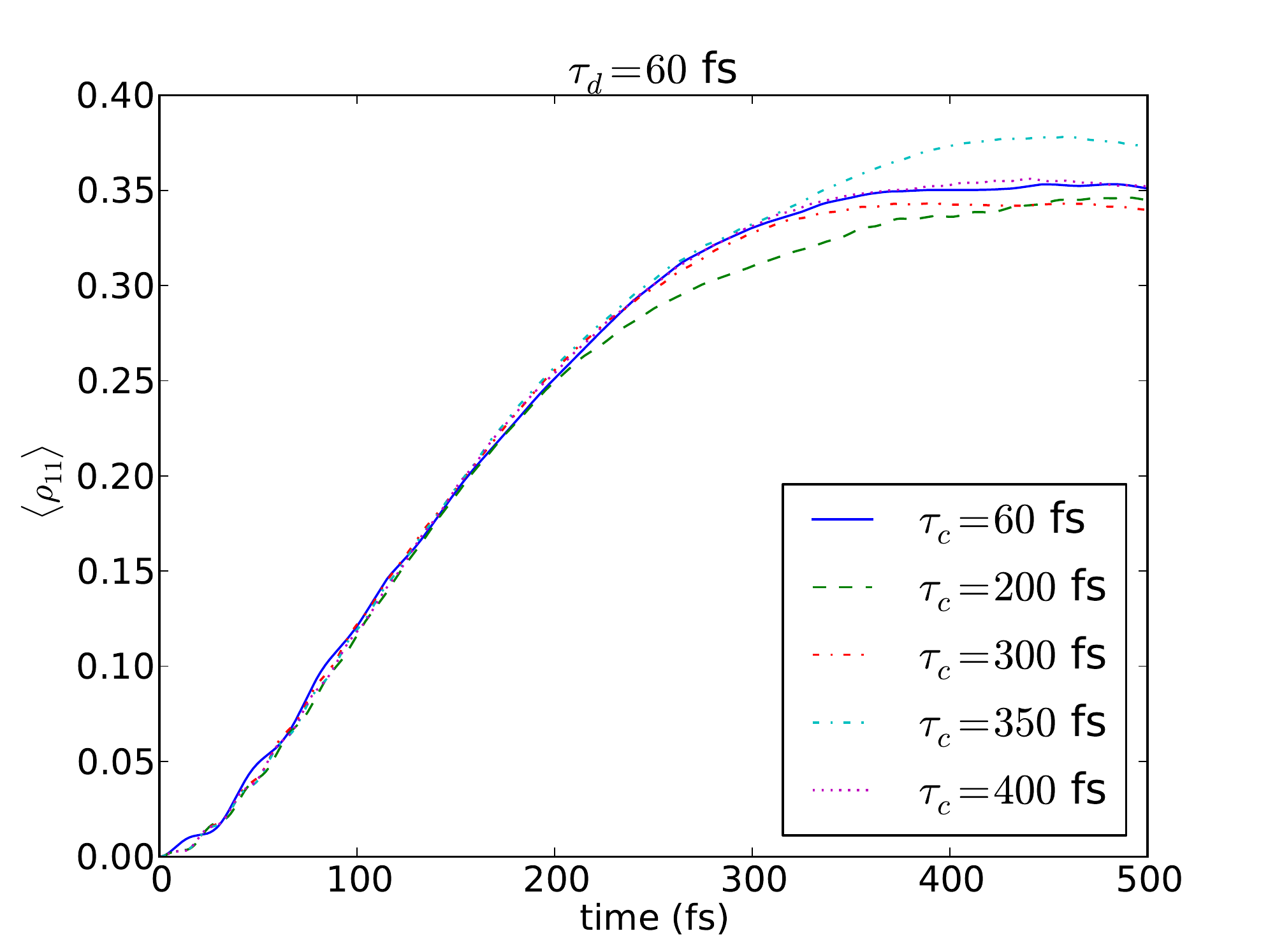}
\includegraphics[scale=0.45]{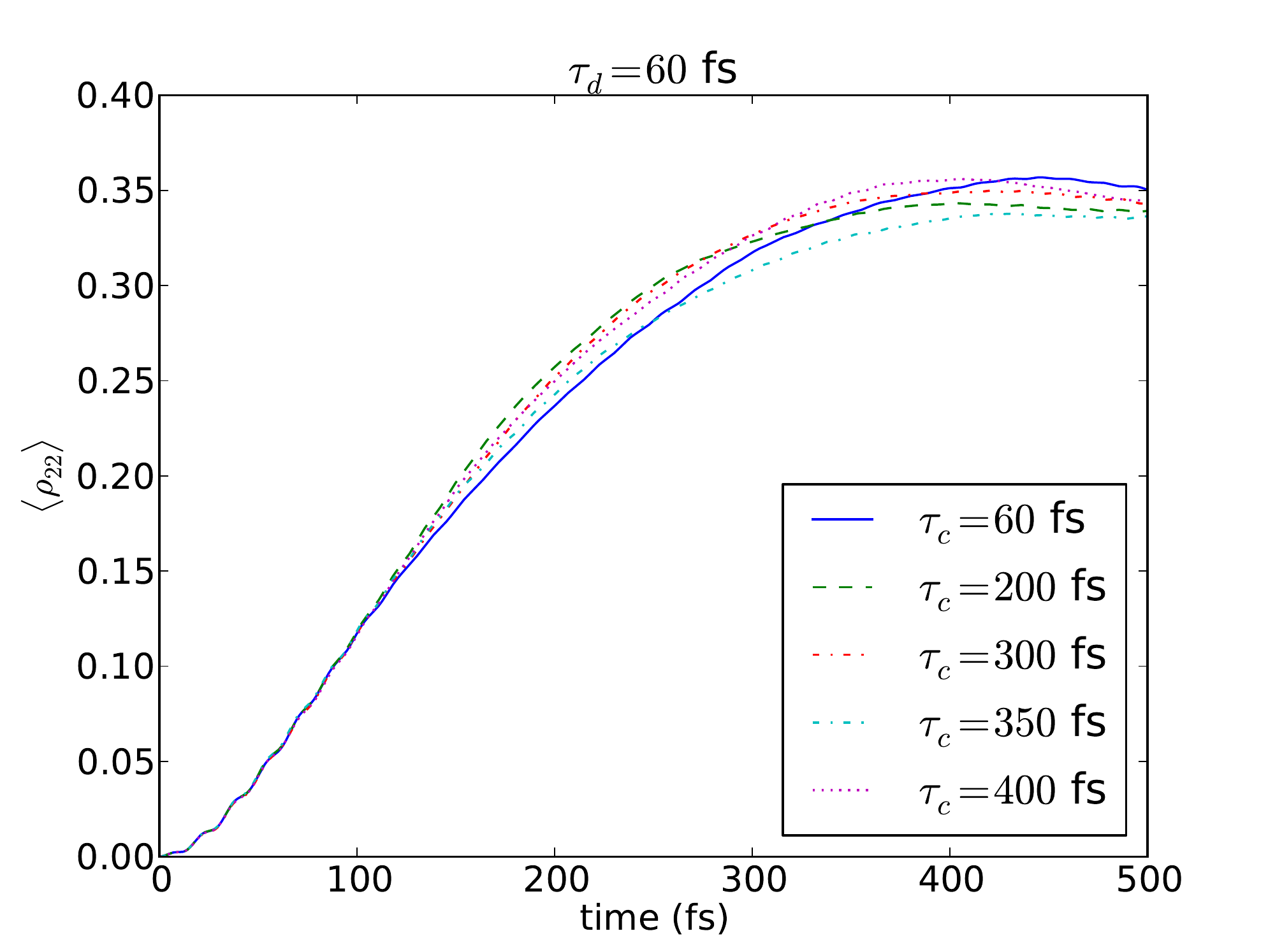}
\caption{(Color online) Populations of excited states $\rho_{11}$ and $\rho_{22}$ for various level spacings, it is clear that they equilibrate to values of $1/3$. The radiation coherence time, $\tau_{d}=60$ fs for all figures.}
\label{cohpop60}
\end{center}
\end{figure}

\begin{figure}[h]
\begin{center}
\includegraphics[scale=0.43]{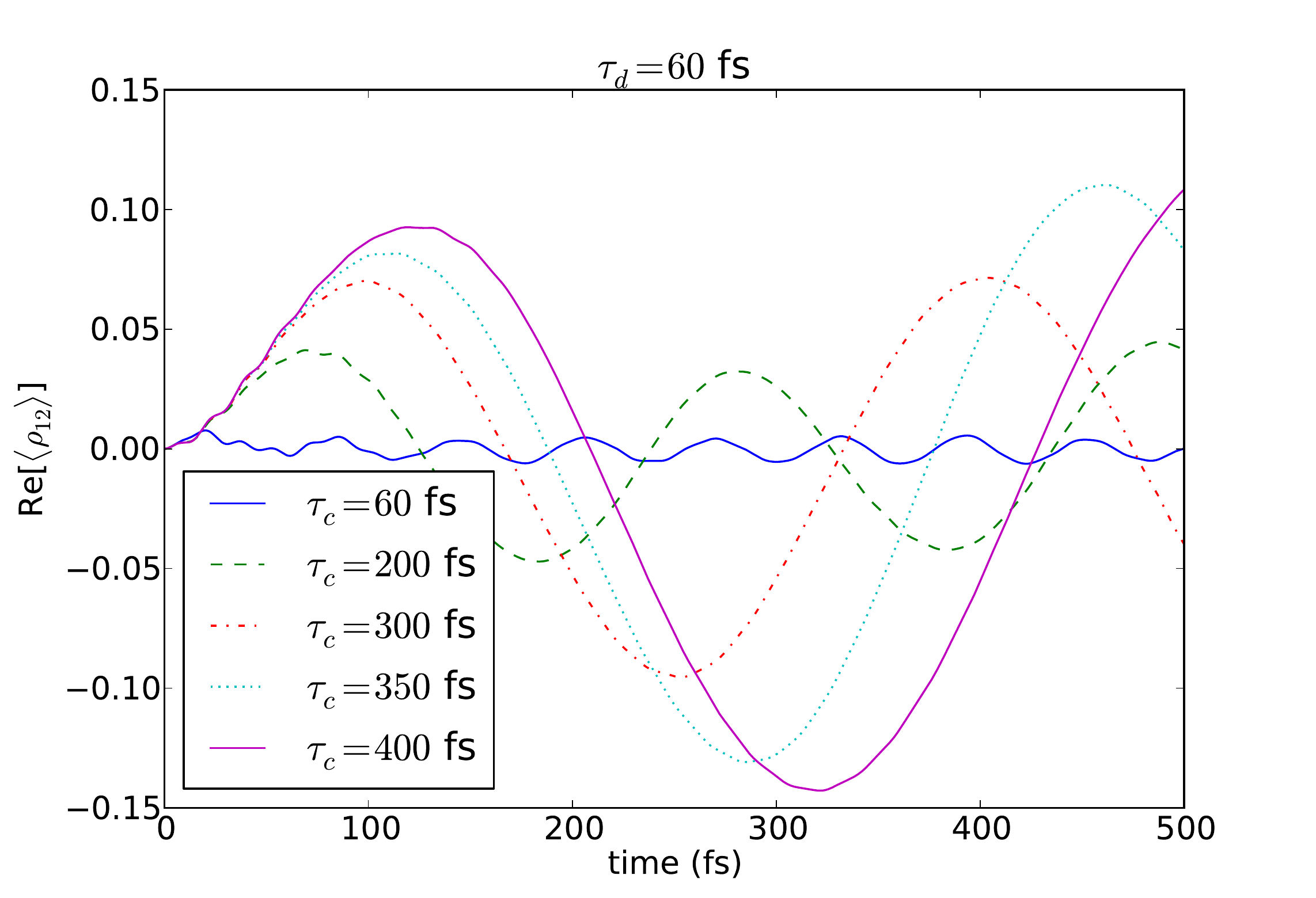}
\caption{(Color online) Excited state coherences $\rho_{12}$ for various level spacings. As level spacing gets smaller the amount of coherence between the states get larger as seen in ref \cite{zaheen}. The radiation coherence time, $\tau_{d}=60$ fs for all figures.}
\label{cohpop601}
\end{center}
\end{figure}

\begin{figure}[b]
\begin{center}
\includegraphics[scale=0.45]{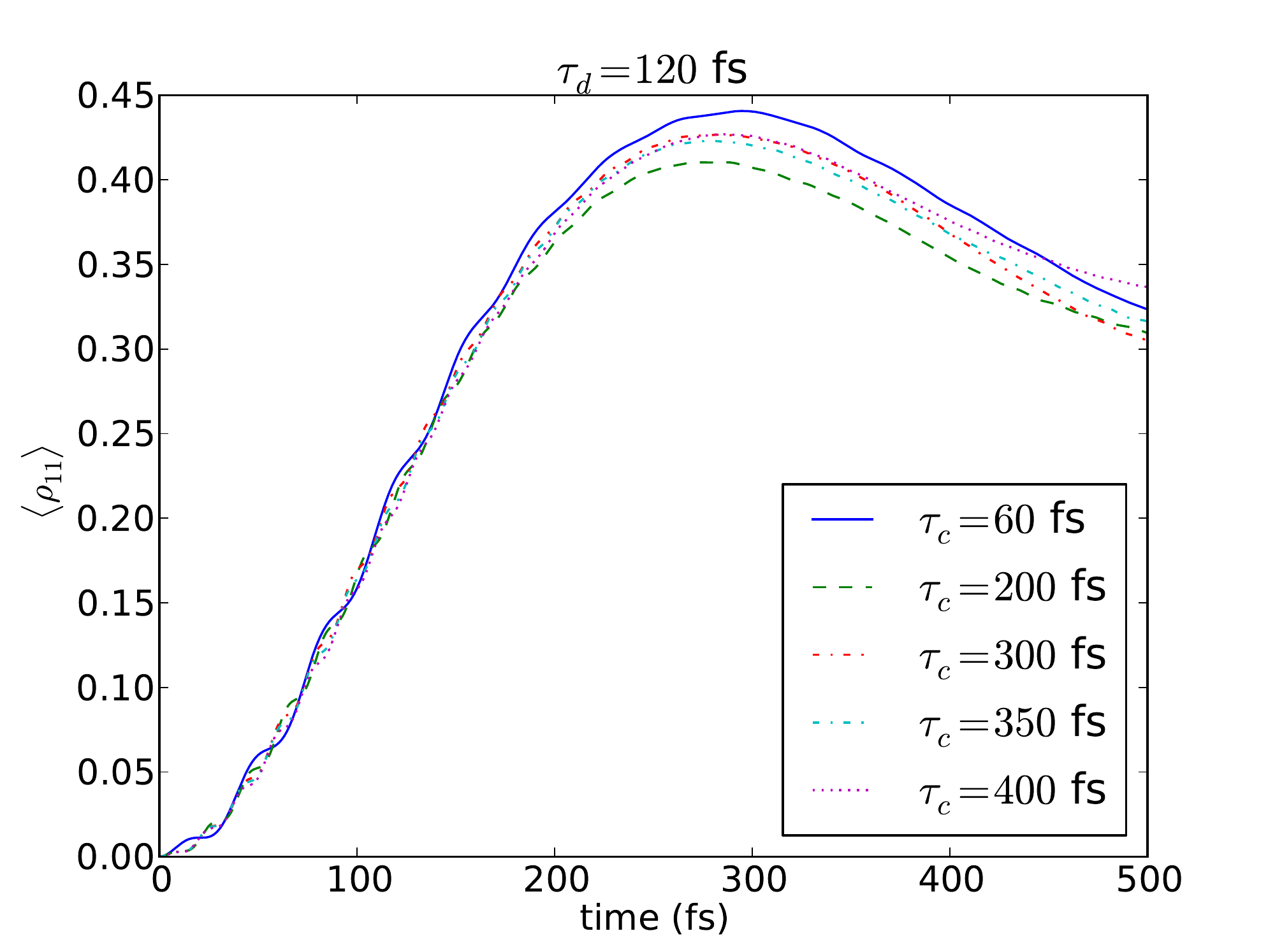}
\includegraphics[scale=0.45]{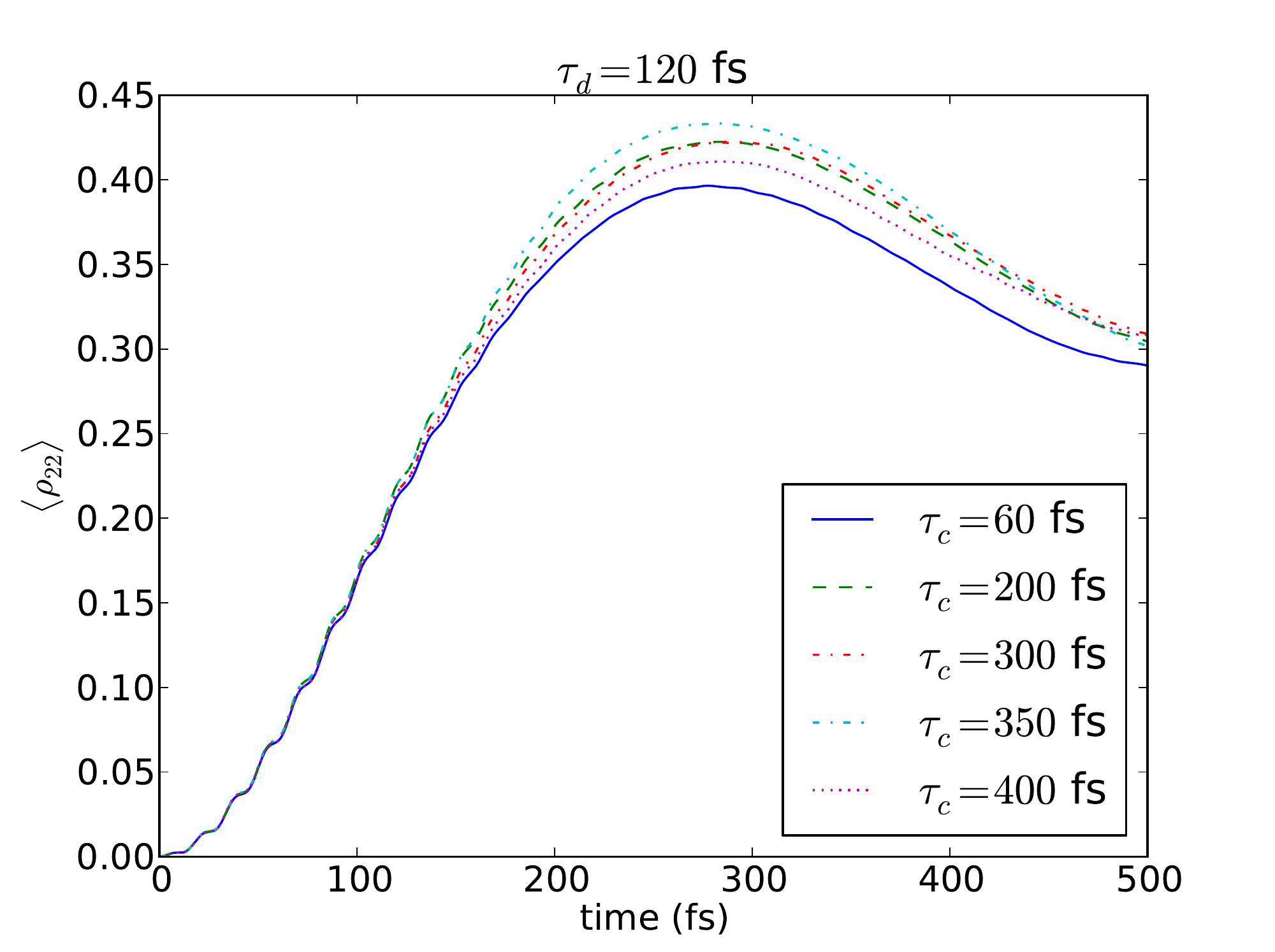}
\caption{(Color online) Populations of excited states $\rho_{11}$ and $\rho_{22}$ for various level spacings, it is clear that they equilibrate to values of $1/3$. The radiation coherence time, $\tau_{d}=120$ fs for all figures.}
\label{cohpop120}
\end{center}
\end{figure}

\begin{figure}[h]
\begin{center}
\includegraphics[scale=0.45]{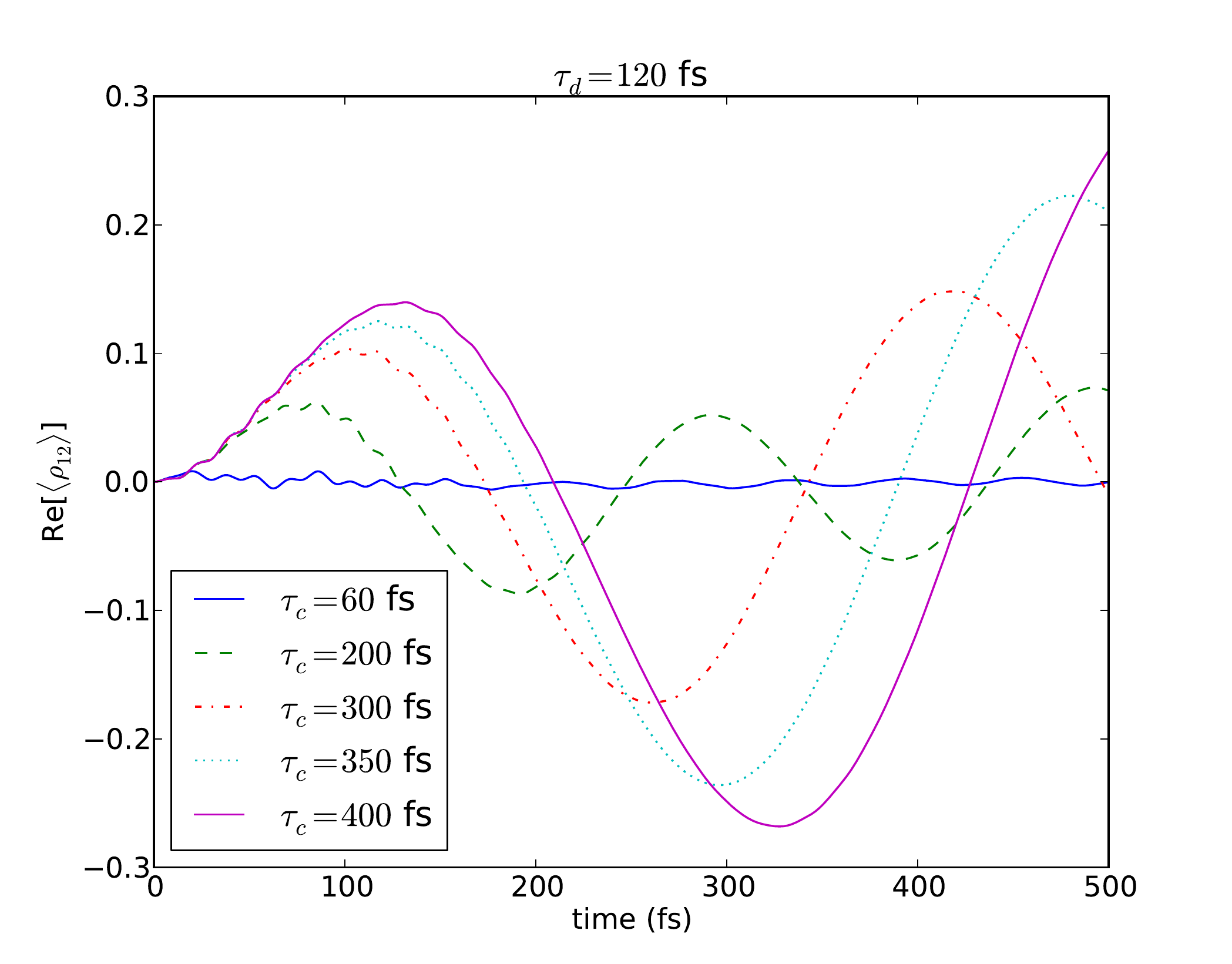}
\caption{(Color online) Excited state coherences $\rho_{12}$ for various level spacings. As level spacing gets smaller the amount of coherence between the states get larger as seen in ref \cite{zaheen}. The radiation coherence time, $\tau_{d}=120$ fs for all figures.}
\label{cohpop1201}
\end{center}
\end{figure}

In this section we outline excitation of a closed but not degenerate $V$ system with two uncorrelated collisionally broadened CW sources tuned to the transition frequencies of the ground to the first two excited states. 

In the previous section it was shown that the use of uncorrelated white noise fields leads to no coherence between the excited states, but instead leads to incoherent pumping to the excited states and eventual equilibrium. However, excitation by two uncorrelated collisionally broadened CW sources it is possible to generate a transient coherent response. 
The Hamiltonian of a closed non degenerate $V$ system irradiated by two uncorrelated collisionally broadened CW sources, in the RWA and EDA is as follows:
\begin{equation}
\mathcal{H}=\left(\begin{array}{ccc}
\hbar\omega_{g} & -\mu_{1}\epsilon_{1}\left(t\right) & -\mu_{2}\epsilon_{2}\left(t\right)\\
-\mu_{1}\epsilon_{1}^{*}\left(t\right) & \hbar\omega_{1} & 0\\
-\mu_{2}\epsilon_{2}^{*}\left(t\right) & 0 & \hbar\omega_{2}
\end{array}\right)
\label{genham}
\end{equation}
In this model, the ground state is designated $|g\rangle$, and two excited states are $|1\rangle$, $|2\rangle$. This describes an interaction between a $V$ level system and two electric fields that couple the following transitions $|g\rangle \rightarrow |1\rangle$ and $|g\rangle \rightarrow |2\rangle$. $\mu_{1}$ is the dipole moment of the $|g\rangle \rightarrow |1\rangle$ transition and $\mu_{2}$ is the dipole moment of the $|g\rangle \rightarrow |2\rangle$ transition. These lasers have a correlation function that is given by Eq. (\ref{corr1}) and each laser's central frequency is tuned to the resonance frequency of the corresponding transition. The statistics of the light, however, are given by:
\begin{equation}
\langle \epsilon^{*}_{i}(t) \epsilon_{j}(t') \rangle = \delta_{ij}\epsilon_{0}^{2}e^{-i\omega_{0}\left(t'-t''\right)}e^{-\frac{|t'-t''|}{\tau_{d}}}
\label{corr3}
\end{equation}
Where $\delta_{ij}$ is the Kronecker delta function. In this scenario, the potential initial self correlation encountered in the schemes outlined in ref \cite{kozlov, scully} is avoided.
We use this Hamiltonian to solve the Liouville-von Neumann equation exactly numerically:
\begin{equation}
\dot{\rho}=\frac{i}{\hbar}\left[\rho,\mathcal{H}\right]
\label{lio}
\end{equation}

The results of our numerics are presented in Figs \ref{wienerres},\ref{wienerres2}. Generation of the collisionally broadened CW laser is covered elsewhere \cite{zaheen} and details on how to force the fields to be uncorrelated are presented in the appendix. Briefly our approach is as follows: stochastic realizations of the noisy field $\{ \epsilon(t) \}$ is generated which is then subsequently used to generate a realization of the system response $\{ \rho(t) \}$. These responses are then collected and ensemble averaged to produce $\langle \rho(t) \rangle$. To ensure the lack of correlation between the two fields we use different random seeds for important parameters such as phase change and phase change time. We calculated the response of a system to radiation with two different coherence times, $\tau_{d} = 60$ fs and $\tau_{d} = 120$ fs. The dipole moments of both $|g\rangle \rightarrow |1\rangle,|2\rangle$ were set to be equal $\mu_{1} = \mu_{2}$. 

Several measures were used to determine the purity and the mixed state character of the system. The coherences as a fraction of excited state population $\mathcal{C} \equiv |\rho_{12}|/(\rho_{11} + \rho_{22})$ was plotted as function of time. The purity of the system, $\text{Tr}[\rho^{2}]$ was also plotted as function of time. These are presented in Figs \ref{wienerres},\ref{wienerres2}. 

The incident field strength was set to $\mu_{i}\epsilon_{0}/\hbar = 10$ THz for all calculations unless otherwise specified. This value of the Rabi frequency was chosen for numerical convenience. Our main results, encapsulated in the measure $\mathcal{C}$, are independent of the Rabi frequency we choose. 

We can see that even though there is no initial self correlation, it is still possible to have a transient coherent response for various excited state splittings using collisionally broadened CW laser excitation. The coherences seen in Figs \ref{wienerres},\ref{wienerres2} are due to the partial coherence of the field for finite times, as demonstrated in Eq. \ref{g1wien}.

An interesting trend develops: the larger the excited state splitting relative to the radiation coherence time, $\tau_{d}$, the more rapid the decay to a mixed state. This can be observed in the purity of the total state as a function of time, as well as the rapid deterioration of excited state coherence fraction. However, as the excited states splitting becomes smaller relative to the radiation coherence time, $\tau_{d}$, the excited state coherences become a larger fraction of excited state population, manifested in the measure $\mathcal{C}$. As $\tau_{d}$ is increased then naturally the response becomes more coherent.

The purity of the total state decreases as a function of time. However, the rate of this decrease becomes smaller as $\tau_{d}$ increases, i.e. the more coherent the field the smaller the decoherence experienced by the system. Purity also decreases at a smaller rate for systems with small excited state splitting. 

For the cases studied, regardless of the excited state splitting, the excited state populations equilibrate to $1/3$, similar to the case of excitation by uncorrelated white light sources, this is demonstrated in Figs \ref{cohpop60},\ref{cohpop120}. However, the amplitude of the excited state coherences increases as excited state splitting becomes smaller, as seen in Figs \ref{cohpop601},\ref{cohpop1201}. From previous investigations of excitation by partially coherent light \cite{zaheen}, it was shown that as the excited state splitting becomes smaller relative to radiation coherence time $\tau_{d}$, the coherences become a larger fraction of the excited state population. This is due to the coherences being inversely proportional to the level spacing $\omega_{ij}$ as demonstrated in ref \cite{zaheen}. A more in depth discussion of the issue is found there. The constraint of the uncorrelated fields forces the populations to equilibrate to a steady value as the coherences become larger. Hence the coherence fraction, $\mathcal{C}$, for small excited state splitting is large.

The purity of the system, $\text{Tr}[\rho^{2}]$ is affected by the populations of the levels $\rho_{gg}$, $\rho_{11}$, and $\rho_{22}$ as well as the excited state coherences $\rho_{12}$ and the ground to excited state correlations. In all the cases that were studied, the populations of the states approach the same value. The differentiating factor is the excited state coherences and the ground to excited state correlations. The amplitude of the excited state coherences also increases as the excited state period becomes larger, thus purity decreases at a slower rate for large excite state periods relative to the radiation coherence time, $\tau_{d}$.

For small splittings, and thus large excited state periods, the excited state subspace remains unusually coherent. This offers an interesting application to quantum optics which requires coherence for phenomena such as EIT or lasing \cite{orszag}. This degree of excited state coherence can remain coherent for a long time (approx. $500$ fs) for certain cases.

\section{Conclusion}

In this paper we examined excitation of a $V$ level system by two uncorrelated white noise and collisionally broadened CW sources. For the case of white noise excitation, we have demonstrated that it is \emph{not} possible to generate even a transient coherent system response.

We have also shown that a $V$ level system excited using two uncorrelated collisionally broadened CW sources allows for the creation of a transient coherent response. We attribute the transient coherence of the material system to the transient coherence of the field \cite{onephoton}. As one alters the excited state splitting, the nature of the coherent response changes. For small excited state periods (relative to the radiation coherence time $\tau_{d}$) the coherences become a small fraction of the excited state population. However, for large excited state periods (relative to the radiation coherence time $\tau_{d}$), an unusually coherent excited state subspace persists where the coherences of the excited remain large relative to the sum of the excited state populations. 


We must emphasize that not only is the aforementioned scenario physically realizable in the laboratory but also generates coherences between excited states using the partial coherence of the field. Our result should be contrasted with literature models such as \cite{kozlov, scully}, which generate coherences from white noise, a source with ill defined coherence properties. These sources exploit a correlation between the field and the system at some initial time $t_{0}$ to generate coherences between the excited states. 


\acknowledgements 
Z.S. would like to thank Y. Khan, Dr. T. Scholak, Dr. T. V. Tscherbul,  R. Dinshaw, S. M. Park, Z. Vernon, S. Foster, L.F. Campitelli, M. M. Gabra and S. Matern for edifying discussions.

\newpage
\begin{widetext}
\newpage
\appendix

\section{\label{app:radgen}Generation of Uncorrelated Radiation}

\begin{figure}[h]
\begin{center}
\includegraphics[scale=0.5]{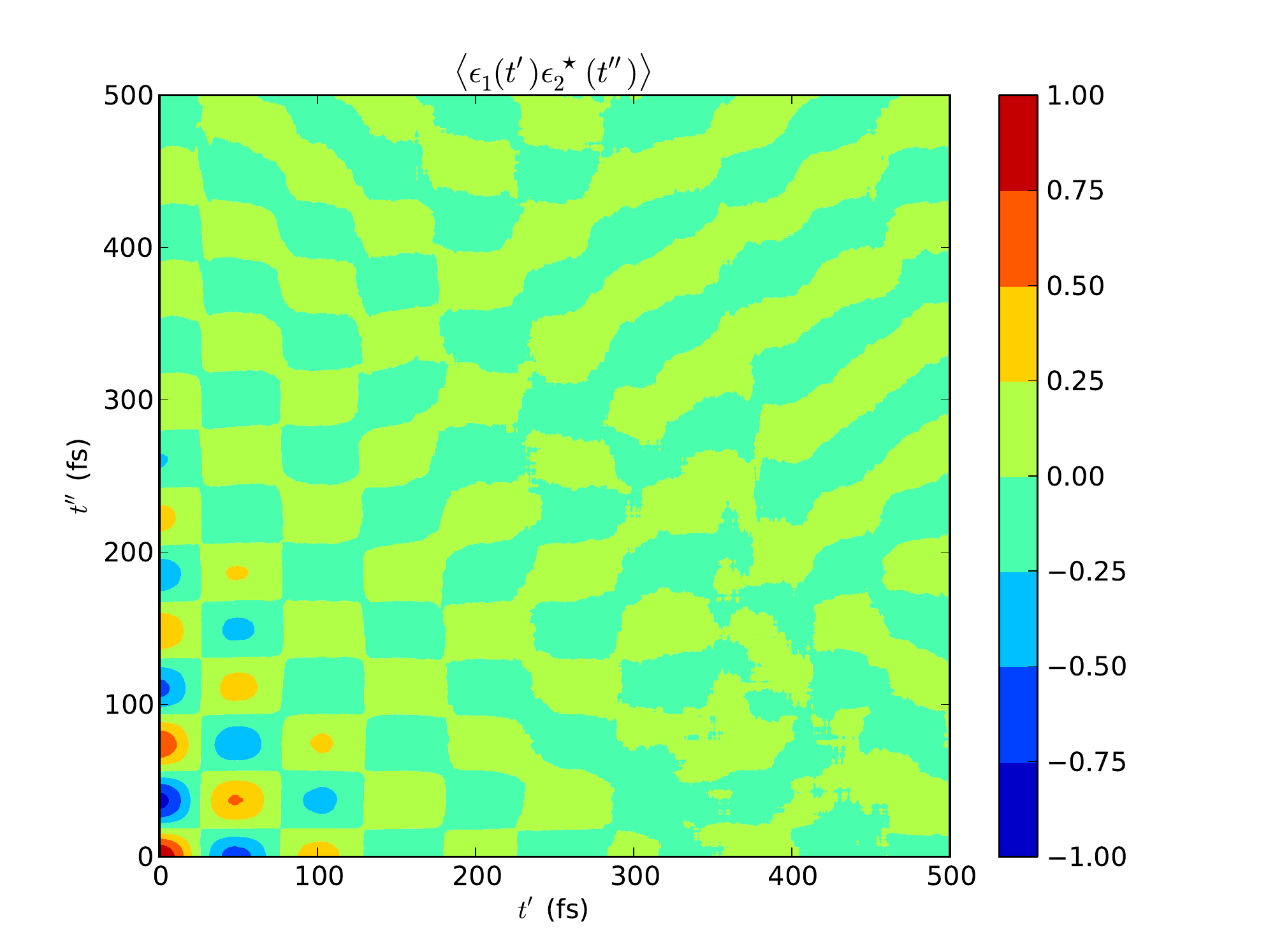}
\caption{(Color online) Correlation function of the two uncorrelated lasers incident on the system. It is clear that these two sources are virtually uncorrelated especially at later times. There is a small initial correlation which disappears at later times.}
\label{uncorrfcn}
\end{center}
\end{figure}

Details on how to numerically generate Wiener noise for the collisionally broadened CW source can be found elsewhere \cite{zaheen}. To generate two uncorrelated fields we use different random seeds for important parameters such as phase change and phase change time. The correlation function of the two uncorrelated fields is plotted in Fig \ref{uncorrfcn}.

A stochastic realization of the noisy pulse $\{ \epsilon(t) \}$ is generated which is then subsequently used to generate a realization of the system response $\{ \rho(t) \}$. These responses are then collected and ensemble averaged to produce $\langle \rho(t) \rangle$. The physical density matrix is represented by this ensemble average $\rho(t) = \langle \rho(t) \rangle$.

\end{widetext}
\end{document}